\documentclass{article}
\usepackage{spconf,amsmath,bm}
\usepackage{comment}
\usepackage{graphicx}
\usepackage{bm}
\usepackage{url}
\usepackage{amssymb}
\usepackage{cite}

\DeclareMathOperator{\clip}{clip}
\title{Stable Training of {DNN} for Speech Enhancement based on Perceptually-Motivated Black-Box Cost Function
}
\name{Masaki Kawanaka$^{\dagger}$,  Yuma Koizumi$^{\ddag}$, Ryoichi Miyazaki$^{\dagger}$, and Kohei Yatabe$^{\star}$}
\address{
$^{\dagger}$National Institute of Technology, Tokuyama College, Yamaguchi, Japan\\  
$^{\ddag}$NTT Media Intelligence Laboratories, Tokyo, Japan\\ 
$^{\star}$Department of Intermedia Art and Science, Waseda University, Tokyo, Japan
\vspace{-7pt}
}
\begin{document}
\ninept
\maketitle
\begin{abstract}
Improving subjective sound quality of enhanced signals is one of the most important missions in speech enhancement.
For evaluating the subjective quality, several methods related to perceptually-motivated objective sound quality assessment (OSQA) have been proposed such as PESQ (perceptual evaluation of speech quality).
However, direct use of such measures for training deep neural network (DNN) is not allowed in most cases because popular OSQAs are non-differentiable with respect to DNN parameters.
Therefore, the previous study has proposed to approximate the score of OSQAs by an auxiliary DNN so that its gradient can be used for training the primary DNN.
One problem with this approach is instability of the training caused by the approximation error of the score.
To overcome this problem, we propose to use stabilization techniques borrowed from reinforcement learning.
The experiments, aimed to increase the score of PESQ as an example, show that the proposed method (i) can stably train a DNN to increase PESQ, (ii) achieved the state-of-the-art PESQ score on a public dataset, and (iii) resulted in better sound quality than conventional methods based on subjective evaluation.
\end{abstract}
\begin{keywords}
Speech enhancement, sound quality assessment, perceptual evaluation of speech quality, function approximation.
\end{keywords}
\section{Introduction}
\label{sec:intro}
Speech enhancement, which aims to recover the target speech from a noisy observed signal, is a fundamental task in a wide range of speech applications including automatic speech recognition~(ASR) \cite{NTTchime} and telecommunication \cite{Kobayashi}.
In these applications, the objectives of enhancement are different; the former is for helping machine listening, while the latter is for helping human listening.
This study focuses on the latter, where the subjective sound quality of the enhanced speech signal is the target of improvement.

Over the last decade, the use of deep neural network (DNN) for speech enhancement has substantially advanced the state-of-the-art performance \cite{Wang_2018,Erdogan_2015,segan,Koizumi_ICASSP_2017,Erdogan_2018_INTERSPEECH,mmsegan,dfl,Koizumi_TASL_2018,metricgan,sergan,Koizumi_icassp_2020,Takeuchi_icassp_2020_Real_time,Takeuchi_icassp_2020_Invertible}.
The popular strategy is to estimate a time-frequency~(T-F) mask by a DNN and apply it in the short-time Fourier transform~(STFT)--domain \cite{Wang_2018}, where the enhanced signal is obtained by the inverse STFT.
Ordinarily, DNNs are trained by back-propagation to minimize a mathematically-defined differentiable cost function such as the mean squared/absolute error \cite{Erdogan_2015} and the signal-to-distortion ratio (SDR) \cite{Erdogan_2018_INTERSPEECH}.
Unfortunately, it has been shown that such mathematically-defined cost functions do not guarantee to improve subjective sound quality \cite{Koizumi_TASL_2018}.
For improving the sound quality of enhanced speech signals, human-perception-based measures for objective sound quality assessments~(OSQA), such as PESQ (perceptual evaluation of speech quality) \cite{PESQ}, has been applied to the training of DNNs.

The difficulty in the training of DNN based on the score of OSQA is its non-differentiable nature which restricts the use of back-propagation.
In the previous studies, two types of strategies have been proposed to circumvent this difficulty \cite{Koizumi_ICASSP_2017,Koizumi_TASL_2018,metricgan}.
Koizumi \textit{et~al.}~formulated the training as a black-box optimization problem and adopted techniques from reinforcement learning (RL) that approximate the gradient using a sampling algorithm \cite{Koizumi_ICASSP_2017,Koizumi_TASL_2018}.
MetricGAN proposed by Fu \textit{et al.} \cite{metricgan} is the other approach which utilizes an auxiliary DNN to approximate the score of OSQA as the generative-adversarial-network (GAN) \cite{gan}.
This function-approximation-based strategy allows to back-propagate the information of OSQA to the primary DNN which enhances the signals.
While these methods effectively improved the sound quality, their problem is instability of the training.
The targeted score of OSQA on the test dataset does not stably increase (see Fig.\,5 of \cite{Koizumi_TASL_2018} and Fig.\,2 of \cite{metricgan}), which can be a cause of failure for some situations.

\begin{figure*}[t]
  \centering
  \centerline{\includegraphics[width=\linewidth]{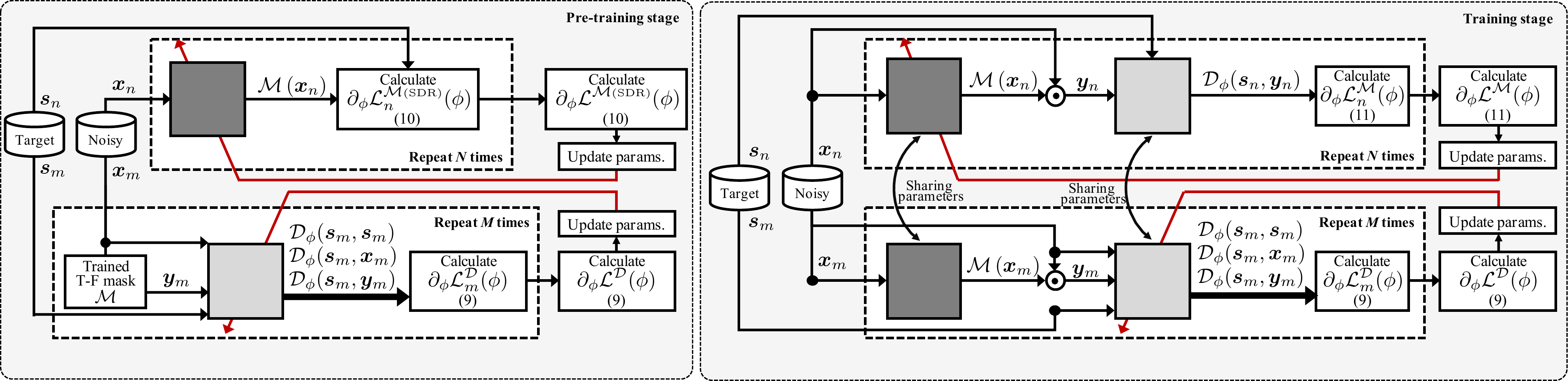}}
  \vspace{-5pt}
\caption{Overview of training procedure of proposed method.}
\label{fig:overview}
\end{figure*}

In this study, we propose the use of stabilization techniques for the function-approximation-based method as shown in Fig.\,\ref{fig:overview}.
For stably training the auxiliary DNN for approximating the score of OSQA, we design a new cost function and adopt training techniques in RL and other machine learning areas.
We conducted experiments training a DNN based on PESQ as an example, and the results show that the proposed method
(i) can stably train a DNN to increase PESQ,  
(ii) achieved the state-of-the-art PESQ score on a public dataset, and
(iii) obtained better sound quality than conventional methods based on subjective evaluation.
\vspace{-2pt}
\section{Conventional Methods}
\label{sec:format}
\subsection{DNN-based Speech Enhancement using T-F mask}
Let $T$-points-long time-domain observation $\bm{x} \in \mathbb{R}^{T}$ be a mixture of a target signal $\bm{s}$ and noise $\bm{n}$ as $\bm{x} = \bm{s} + \bm{n}$.
The goal of speech enhancement is to recover $\bm{s}$ from $\bm{x}$, where the use of DNN has substantially advanced the state-of-the-art performance.
A popular strategy is to use a DNN for estimating a T-F mask in the STFT-domain.
Let $\mathcal{F} : \mathbb{R}^{T} \to \mathbb{C}^{F \times K}$ be the STFT where $F$ and $K$ are the number of frequency and time bins, respectively.
A general form of DNN-based speech enhancement using a T-F mask can be written as
\begin{align}
\bm{y} = 
\mathcal{F}^{\dag}\left(
\mathcal{M}_{\theta\!}\left(
\bm{x} 
\right) \odot
\mathcal{F}\left(
\bm{x}
\right)
\right),
\label{eq:mask}
\end{align}
where 
$\bm{y}$ is the estimate of $\bm{s}$, $\mathcal{F}^{\dag}$ is the inverse-STFT, $\odot$ is the element-wise product, and $\mathcal{M}$ is a DNN for estimating the T-F mask.
The set of parameters of the DNN $\theta$ is trained to minimize a cost function $\mathcal{L}(\bm{s}, \bm{y})$ by iterating the gradient descending procedure:
\begin{align}
\theta^{t+1} \gets \theta^{t} - \lambda\, \partial_{\theta} 
\mathcal{L} (\bm{s}, \bm{y}),
\label{eq:grad_method}
\end{align}
where $\lambda>0$, and $\partial_{\theta}$ is the differential operator w.r.t.~$\theta$.

To apply a gradient-descent-type algorithm as in \eqref{eq:grad_method}, derivatives of both $\mathcal{L}$ and $\mathcal{M}$ are required for computing $\partial_{\theta} \mathcal{L} (\bm{s}, \bm{y})$.
Since $\partial_{\theta} \mathcal{M}$ is usually computed by the back-propagation, differentiability of the cost function $\mathcal{L}$ is the matter for the algorithm.
The mean squared/absolute error \cite{Erdogan_2015} and SDR \cite{Erdogan_2018_INTERSPEECH} are some examples of differential cost functions popular in speech enhancement.
However, these mathematically-defined cost functions do not guarantee to improve the subjective sound quality of the enhanced signals \cite{Koizumi_TASL_2018} because they do not take the perceptual concepts into account.

\subsection{OSQA-based Cost Function for Speech Enhancement}
Instead of such mathematically-defined differentiable cost functions, some perceptually-motivated functions such as PESQ have been considered in DNN-based speech enhancement.
Let $\mathcal{P}(\bm{s}, \bm{y})$ be an OSQA score evaluated between $\bm{s}$ and $\bm{y}$.
To incorporate it into training of DNN, two approches have been proposed  \cite{Koizumi_ICASSP_2017,Koizumi_TASL_2018,metricgan}.

The first approach \cite{Koizumi_TASL_2018} considered the expectation w.r.t.~$\bm{x}, \bm{y}$,
\begin{equation}
\mathcal{L}^{\mathcal{P}} = -\mathbb{E} \left[ \mathcal{P}(\bm{s}, \bm{y})
\right]_{\bm{x}, \bm{y}},
\label{eq:osqa_cost}
\end{equation}
as the cost function.
To calculate $\partial_{\theta} \mathcal{L}^{\mathcal{P}} (\bm{s},\bm{y})$, Koizumi \textit{et al.} \cite{Koizumi_TASL_2018} considered a sampling algorithm used in RL.
They rewrote (\ref{eq:osqa_cost}) as
\begin{equation}
\mathcal{L}^{\mathcal{P}} = 
\int p(\bm{x})
\int
\mathcal{P}(\bm{s}, \bm{y})\,
q_{\theta} (\bm{y} | \bm{x})\,
d\bm{y} d\bm{x},
\end{equation}
where $q_{\theta} (\bm{y} | \bm{x})$ is a conditional distribution of $\bm{y}$ given $\bm{x}$.
As the goal is to train a DNN for recovering $\bm{y}$, $q_{\theta}$ is considered to consist of a DNN.
Then, by using the log-derivative trick, $\partial_{\theta} \mathcal{L}^{\mathcal{P}}$ is given by
\begin{equation}
\partial_{\theta}
\mathcal{L}^{\mathcal{P}} = 
\mathbb{E} \left[
\mathcal{P}(\bm{s}, \bm{y})\,
\partial_{\theta} \ln q_{\theta} (\bm{y} | \bm{x})
\right]_{\bm{x}, \bm{y}}.
\label{eq:RL_diff}
\end{equation}
When $\ln q_{\theta} (\bm{y} | \bm{x})$ is differentiable w.r.t $\theta$ and $\bm{y}$ can be drawn from $q_{\theta} (\bm{y} | \bm{x})$,
(\ref{eq:RL_diff}) can be approximately calculated.
The problem of this approach is that training takes a long-time for stabilizing it.
This is because the expectation in (\ref{eq:RL_diff}) is numerically calculated by the Monte Carlo method, and stabilization of the random-sampling-based expectation requires a huge number of samples.

The second approach, MetricGAN proposed by Fu {\it et al.} \cite{metricgan}, is based on function approximation of $\mathcal{P}(\bm{s}, \bm{y})$.
In this method, the score of OSQA is approximated by using an auxiliary DNN $\mathcal{D}$ as
\begin{align}
\mathcal{P}(\bm{s}, \bm{y})
\approx \mathcal{D}_{\phi}(\bm{s}, \bm{y}),
\end{align}
where $\phi$ is the set of parameters of $\mathcal{D}$.
Since $\mathcal{D}$ is differentiable w.r.t. $\bm{y}$,
$\partial_{\theta} \mathcal{D}_{\phi}(\bm{s}, \bm{y})$ can be calculated via back-propagation.
In this method, $\mathcal{D}$ and $\mathcal{M}$ are trained alternately.
First, $\mathcal{D}$ is updated to decrease the following cost function:
\begin{align}
\mathcal{L}^{\mathcal{D}{\mbox{\tiny (GAN)}}}
=
\mathcal{E}_{\phi}( \bm{s} )
+ 
\mathcal{E}_{\phi}( \bm{y} )
, 
\label{eq:Metriclossd}
\end{align}
where
$\mathcal{E}_{\phi}( \cdot )$
is the mean-squared error (MSE) between true and estimated OSQA scores,
$\mathcal{E}_{\phi}( \cdot ) = (
\mathcal{P}(\bm{s}, \cdot) - \mathcal{D}_{\phi}(\bm{s}, \cdot)
)^2$.
Then, 
to update $\mathcal{M}$ so as to obtain the best score for all $\bm{x}$,
the cost function,
\begin{align}
\mathcal{L}^{\mathcal{M}{\mbox{\tiny (GAN)}}} =
\left( 
\mathcal{D}_{\phi}(\bm{s}, \bm{y}) - 1
\right)^2
,
\label{eq:MmetricGAN}
\end{align}
is minimized, where the OSQA score is assumed to be normalized as
$0 \le \mathcal{P}(\bm{s}, \bm{y}) \le 1$,
and therefore
$\mathcal{P}(\bm{s}, \bm{s}) = 1$

It is known that training of GAN is difficult and unstable.
Thus, Fu {\it et al.}~introduced several techniques, including the Spectral Normalization \cite{specnorm}, to stabilize the training of MetricGAN.
However, as can be seen in Fig.\,2 of their paper \cite{metricgan}, its training is still unstable in the early stage of training (at around 20--50 iterations).
\section{Proposed method}
\label{sec:proposed}
Recently, several pieces of literature have reported that, based on the relevance between GAN and RL, training of GAN can be stabilized by adopting techniques of RL \cite{gan_ac}.
In addition, several techniques for stabilizing DNN training have also been proposed in other areas \cite{gan_ac,adam,sgd_better,ac_lr,large_minibatch}.
Therefore, we consider that the training of MetricGAN, or the function-approximation-based approach, can be stabilized by adopting such techniques.
In this section, we describe the techniques that succeeded to stabilize the training.
\subsection{Techniques for Stabilizing DNN Training}
\begin{figure}[t]
  \centering
  \centerline{\includegraphics[width=\linewidth]{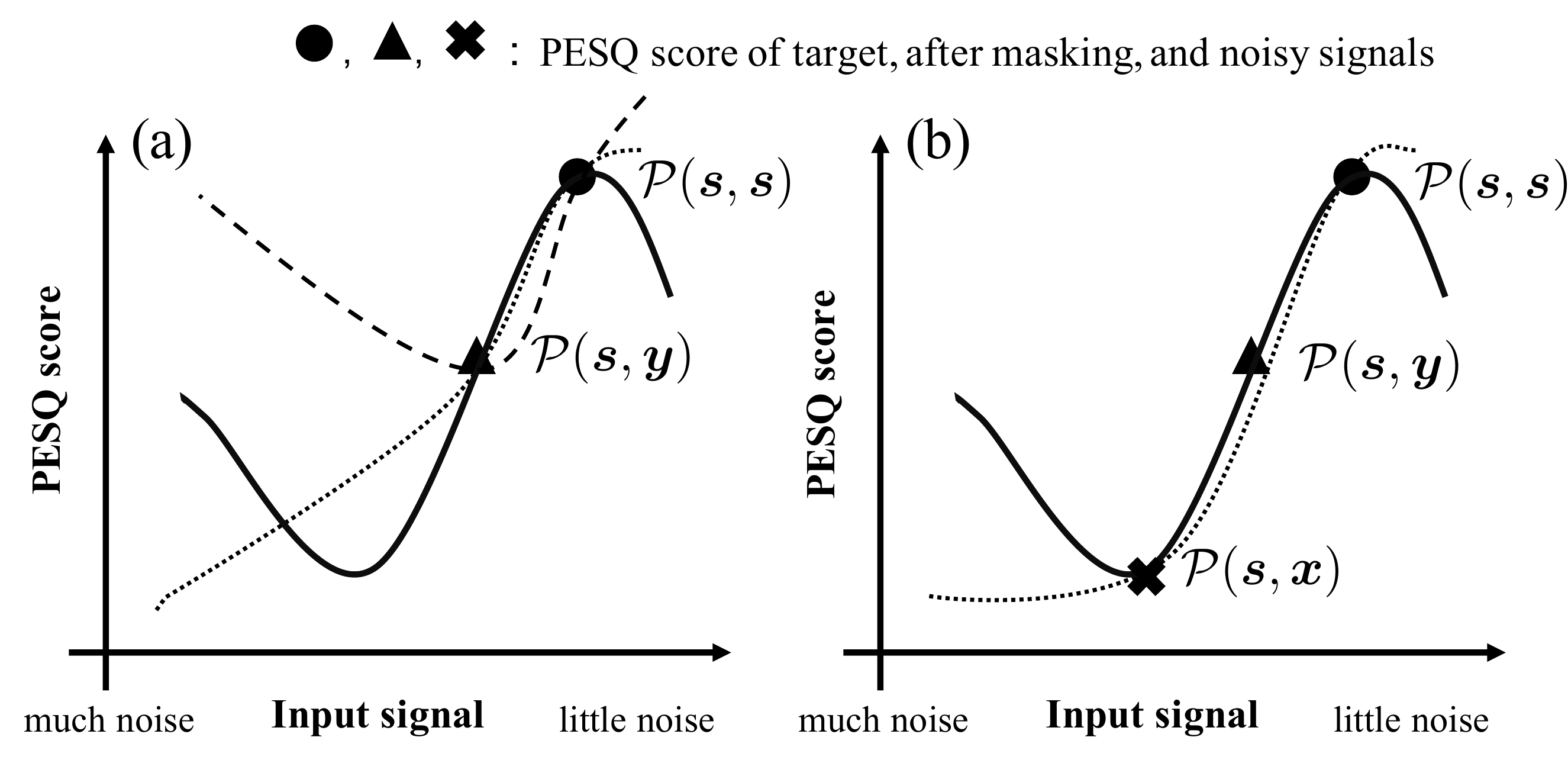}}
    \vspace{-5pt}
\caption{Illustration of difference of OSQA approximation owing to cost functions, (a) MetricGAN $\mathcal{L}^{\mathcal{D}{\mbox{\tiny (GAN)}}}$, and (b) Ours $\mathcal{L}^{\mathcal{D}}$.
The solid-lines represent the true OSQA function, while dotted- and dashed-lines are its approximation by $\mathcal{D}$. Since MetricGAN trains $\mathcal{D}$ using $\mathcal{E}_{\phi}( \bm{s} )$ and $\mathcal{E}_{\phi}( \bm{y} )$ only, $\mathcal{D}$ can become both dotted- and dashed-line in (a). To inform OSQA score of noisy signal, our cost function additionally uses $\mathcal{E}_{\phi}( \bm{x} )$ so that $\mathcal{D}$ becomes dotted-line in (b).
}
\label{fig:cri}
\end{figure}
\subsubsection{Cost function for OSQA score approximation}
\label{sec:cost_D}
First, we consider a cost function better than $\mathcal{L}^{\mathcal{D}{\mbox{\tiny (GAN)}}}$.
Since $\mathcal{L}^{\mathcal{D}{\mbox{\tiny (GAN)}}}$ consists of $\mathcal{E}_{\phi}( \bm{s} )$ and $\mathcal{E}_{\phi}( \bm{y} )$, $\mathcal{D}$ can know the OSQA scores of clean and current-output signals $\bm{s},\bm{y}$ only, i.e., it cannot know the score of noisy signal $\bm{x}$.
Then, it is difficult to approximate the score for noisier signals as illustrated in Fig.\,\ref{fig:cri}-(a).
Such lack of information on $\mathcal{P}(\bm{s}, \bm{x})$ should be the cause of the instability of training.
Therefore, we additionally supervise the OSQA score of a noisy signal as
\begin{align}
\mathcal{L}^{\mathcal{D}}
=
\sum_{m=1}^M
\mathcal{E}_{\phi}( \bm{s}_m ) +
\mathcal{E}_{\phi}( \bm{x}_m ) +
\mathcal{E}_{\phi}( \bm{y}_m )
,
\label{eq:OurlossD}
\end{align}
where $M$ is the minibatch-size of $\mathcal{D}$'s training, and
$\bm{s}_m$,
$\bm{x}_m$, and
$\bm{y}_m$ are 
the $m$th samples of clean, noisy, and output signal in the minibatch, respectively.
Since three points (clean, current, and noisy) of the score are informed to $\mathcal{D}$, we can expect that $\mathcal{D}$ can learn better as in Fig.\,\ref{fig:cri}-(b), and $\mathcal{M}$ can know which is a lower-quality signal.
\subsubsection{Training techniques}
\label{sec:training_tech}
We also adopted some techniques in the training procedures.
Here, we provide a recipe including a pre-training method, optimizer and minibatch-size selection.

\vspace{2pt plus 2pt minus 0.5pt}
(i) {\bf Pre-training:}
Training of $\mathcal{M}$ based on a differentiable cost function is easier than that of using OSQA scores. 
Although SDR does not reflect the subjective sound quality, signals with higher SDR tends to result in higher OSQA score. 
Therefore, pre-training of $\mathcal{M}$ using SDR should be effective.
Thus, before training by OSQA score, we train $\mathcal{M}$ using the SDR-based cost function \cite{Erdogan_2018_INTERSPEECH} defined as
\begin{align}
\mathcal{L}^{\mathcal{M}{\mbox{\tiny (SDR)}}} = 
\sum_{n=1}^N
\clip_{\alpha}
\left[
10 \log_{10}
\frac{ \lVert \bm{s}_n \rVert_2^2 }{ \lVert \bm{s}_n - \bm{y}_n \rVert_2^2 }
\right],
\label{eq:sdr}
\end{align}
where 
$\clip_{\alpha}[x] = \alpha \cdot \tanh (x / \alpha)$,
$\alpha = 20$ is a clipping parameter,
$N$ is the minibatch-size of $\mathcal{M}$'s training, and
$\bm{s}_n$ and $\bm{y}_n$ are 
the $n$th samples of clean and output signals in the minibatch, respectively.
After the pre-training of $\mathcal{M}$ using \eqref{eq:sdr}, $\mathcal{D}$ is also pre-trained using the cost function \eqref{eq:OurlossD} with fixed $\mathcal{M}$.
In the pre-training stage, the minibacth-size was set to 5 for both networks.

\vspace{2pt plus 2pt minus 0.5pt}
(ii) {\bf Optimizer:}
In MetricGAN, the adaptive moment estimation (Adam) \cite{adam} was used as the optimizer for both $\mathcal{M}$ and $\mathcal{D}$.
However, it is known that the Adam optimizer may not improve the generalization performance compared to the stochastic gradient descent (SGD) \cite{sgd_better}.
Therefore, we use the SGD optimizer instead of the Adam optimizer in the training stage. 

\vspace{2pt plus 2pt minus 0.5pt}
(iii) {\bf Minibatch-size:}
Since calculation of OSQA takes long time, using a large minibatch-size is not practical.
However, a small minibatch-size may degrade the approximation accuracy of the OSQA scores because the gradients become less accurate that may results in unstable training.
We experimentally found that the minibatch-size $M = 10$ is the smallest size which stabilizes the training of $\mathcal{D}$ in our experiment.
Based on this finding, we use minibatch-size $N=5$ for $\mathcal{M}$'s training.
The reason for this selection is based on the literature of the actor-critic algorithm in RL which is an optimization method for the function-approximation-based cost function.
In the literature, it is known that the learning rates of the actor (or $\mathcal{M}$) and critic (or $\mathcal{D}$) should be $2:1$ \cite{ac_lr}.
In addition, recent research has reported that decreasing the learning rate has the same effect as increasing minibatch-size in some situations \cite{large_minibatch}, i.e., the learning rate and minibatch-size are related inversely in these context.
Thus, we set the minibatch-sizes to $1:2$ ($N=5$ and $M = 10$) based on the inverse relation.
\subsection{Implementation}
The proposed training procedure is summarized in Fig.\,\ref{fig:overview}.
First, $\mathcal{M}$ is pre-trained using the SDR-based function (\ref{eq:sdr}), and then $\mathcal{D}$ is also pre-trained based on the fixed $\mathcal{M}$ as in the left-hand side of Fig.\,\ref{fig:overview}.
Next, we train $\mathcal{M}$ and $\mathcal{D}$ alternately as follows.
The parameter $\phi$ of $\mathcal{D}$ is updated ten times for decreasing (\ref{eq:OurlossD}).
Then, the parameter $\theta$ of $\mathcal{M}$ is updated twenty times for decreasing the following cost function
\begin{align}
\mathcal{L}^{\mathcal{M}} = -
\sum_{n=1}^N
\mathcal{D}_{\phi}(\bm{s}_n, \bm{y}_n).
\end{align}
Note that, while updating $\mathcal{M}$, $\mathcal{D}$ is fixed, and vice versa.
\section{Experiments}
\label{sec:typestyle}
We conducted three experiments:
(i) a verification experiment for investigating the stabilization effect of the proposed method, 
(ii) an objective experiment using a public dataset, and 
(iii) a subjective experiment.
In all experiments, we utilized the VoiceBank-DEMAND dataset constructed by Valentini {\it et al.} \cite{dataset}
which is openly available and frequently used in the literature of DNN-based speech enhancement \cite{segan,mmsegan,dfl,sergan,metricgan}.
The train and test sets consists of 28 and 2 speakers (11572 and 824 utterances), respectively, and all signals were downsampled to 16 kHz.
As an example of OSQA to be approximated by the DNN, PESQ was considered in this section.
\subsection{Experimental Setups}
The DNN for estimating the T-F mask, $\mathcal{M}$, consisted of two 2-D convolutional neural networks (CNNs) followed by a 1x1 CNN, two linear layers, and two bidirectional long short-term memory (BLSTM)--layers.
This setup is a standard architecture in DNN-based speech enhancement \cite{Erdogan_2018_INTERSPEECH}. 
The input of the DNN was log-amplitude spectrogram of the observed signal $\bm{x}$ whose size was $F \times K$.
The kernel size, stride, and padding of both 2-D CNNs were (5,15), (1,1) and (2,7), respectively.
The number of output channels of the first and second 2-D CNNs were 30 and 60, respectively.
Then, the number of channels was decreased to 1 by the 1x1 CNN.
The dimension of the 1x1 CNN's output $(F \times K)$ was changed by the first linear layer to $D \times K$.
It was passed to the BLSTM layers, and its forward and backward outputs were concatenated so that the output size of this block was $2D \times K$. 
It was converted to $2F \times K$ by the last linear layer.
Finally, the output was split into two $F \times K$ matrices which were used as the real- and imaginary-parts of the  complex-valued T-F mask.
The spectrogram $\mathcal{F} (\bm{x} )$ was multiplied by the estimated complex T-F mask and transformed back to the time-domain as (\ref{eq:mask}), where the STFT parameters, frame shift and window size $(=\text{DFT size})$, were set to 128- and 512-samples, respectively, with the Hann window.
The DNN for approximating PESQ, $\mathcal{D}$, was the same network used in MetricGAN.

In the pre-training stage, $\mathcal{D}$ and $\mathcal{M}$ were trained 200 and 290 epochs, where each epoch included randomly selected 1,000 utterances.
We fixed the learning rate for the initial 100 epochs and then decreased it linearly down to a factor of 100 using Adam optimizer, where we started with a learning rate of 0.001. 
In the training stage, SGD was used as the optimizer and the learning rate was set to 0.001, and it was concluded after 2,000 updates.

\vspace{2pt}
\subsection{Objective experiments}
\begin{figure}[t]
  \centering
  \centerline{\includegraphics[width=\linewidth]{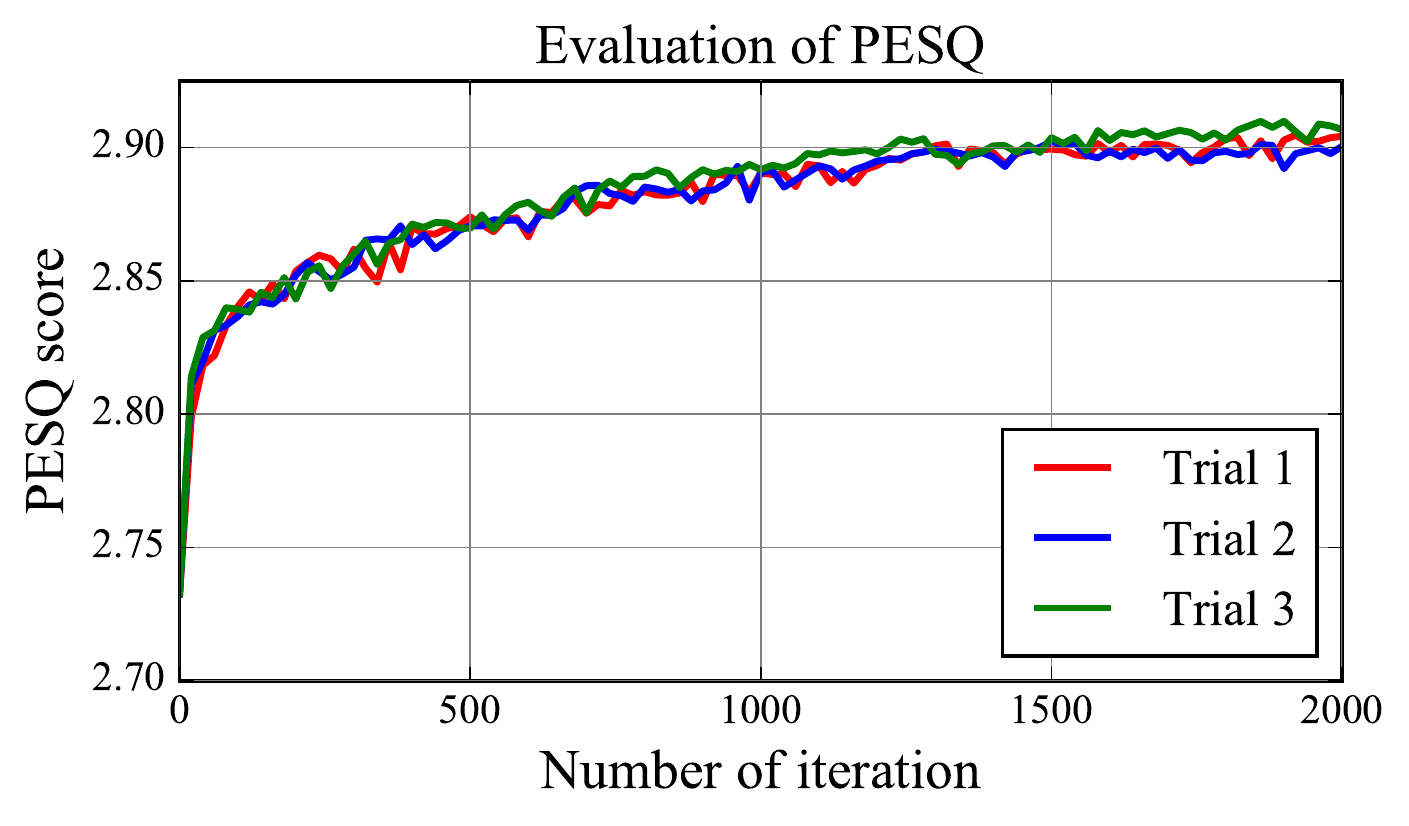}}
    \vspace{-5pt}
\caption{Relationship between number of iterations and PESQ score on test-dataset. Trial1, Trial2, and Trial3 represent the different seed values for randomly choosing training-minibatch in training stage.
}
\label{fig:ite_pesq}
\end{figure}

For the objective evaluation, we conducted two experiments.
First, we conducted an experiment for verifying whether the PESQ score of the test-dataset was stably improved by increasing the number of iterations. 
Figure~\ref{fig:ite_pesq} shows the relationship between the number of iterations and the PESQ score test-dataset, where each minibatch in training stage was randomly chosen by using a seed value.
From Fig.\,\ref{fig:ite_pesq}, the PESQ score was improved stably by increasing the number of iterations for all seed values used for choosing the minibatches in traning stage.
This result indicates that the proposed method succeeded to stabilize the training of DNN-based speech enhancement for increasing the OSQA score.

\begin{table}
\caption{Results of objective evaluation}
\label{tbl:obj}
\begin{center}
\small 
\begin{tabular}{l|cccc} \hline
    Method              & PESQ & CSIG & CBAK & COVL \\\hline 
    Noisy               &1.97 & 3.35 & 2.44 & 2.63 \\ 
    SEGAN \cite{segan} 			&2.16 & 3.48 & 2.94 & 2.80 \\ 
    MMSE-GAN \cite{mmsegan}  		&2.53 & 3.80 & 3.12 & 3.14 \\
    DF-Loss \cite{dfl} 	        &  -  & 3.86 & {\bf 3.33} & 3.22 \\ 
    SERGAN \cite{sergan} 		    &2.62 &    -    &     -   & -      \\ 
    MetricGAN \cite{metricgan}		&2.86 & {\bf 3.99} & 3.18 & {\bf 3.42} \\ \hline
    Pre-train 	    	&2.73 & 3.73 & 2.55 & 3.20 \\ 
    Ours         		&{\bf 2.93} & 3.72 & 2.64 & 3.29 \\ \hline 
  \end{tabular}
\vspace{-10pt}   
\end{center}
\end{table}

\begin{figure}[t]
  \centering
  \centerline{\includegraphics[width=\linewidth]{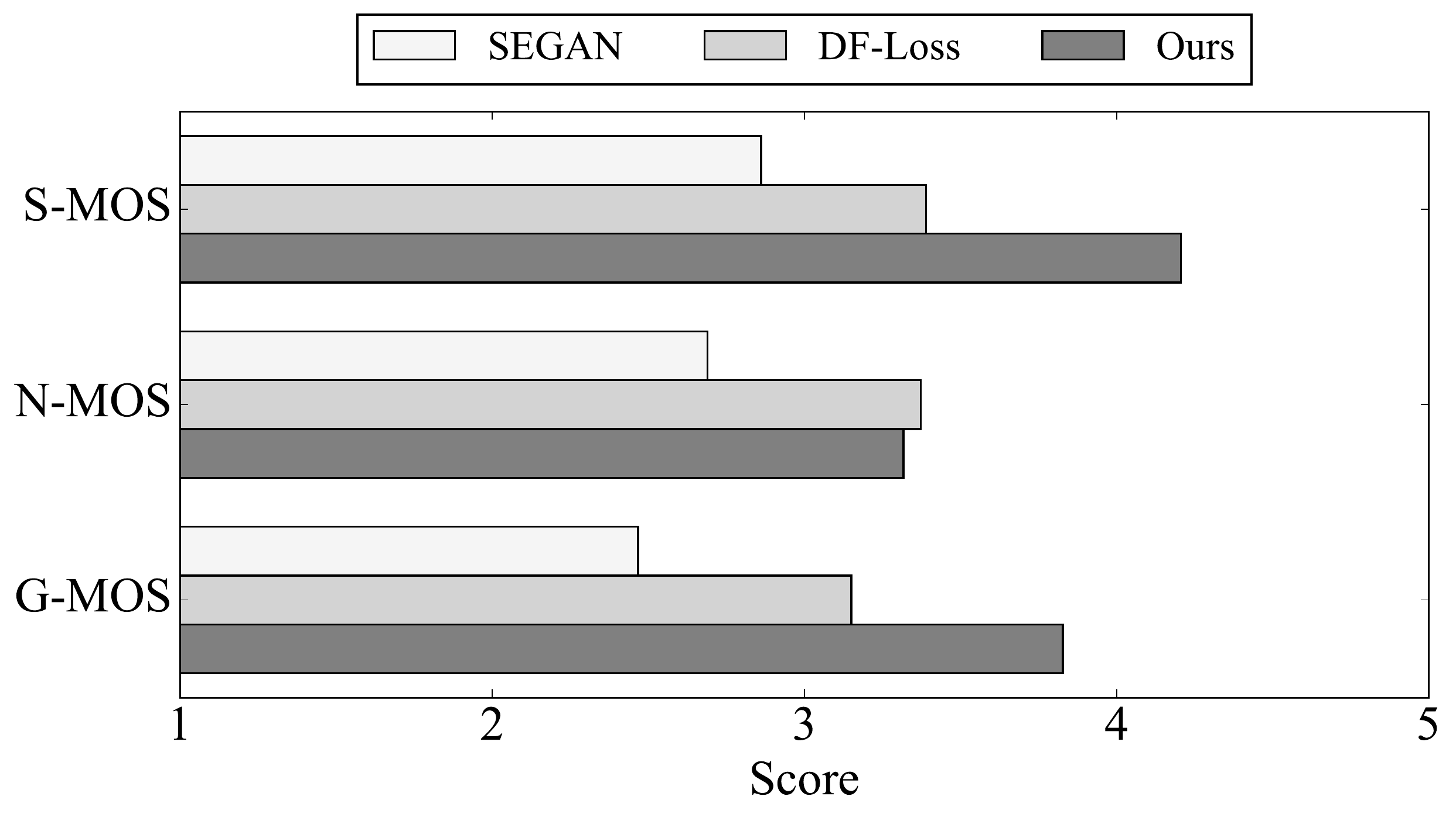}}
    \vspace{-5pt}
\caption{Result of subjective evaluation.}
\label{fig:sub}
\end{figure}

Next, we compared the proposed method with the conventional methods on the same dataset and metrics, where CSIG, CBAK, and COVL are the popular predictor of the mean opinion score (MOS) of the target signal distortion, background noise interference, and overall speech quality, respectively~\cite{eval_metric}.
In this evaluation, we considered SEGAN~\cite{segan}, MMSE-GAN~\cite{mmsegan}, Deep Feature Loss (DF-Loss)~\cite{dfl}, SERGAN~\cite{sergan} and MetricGAN~\cite{metricgan} as the reference of conventional methods because these methods have been evaluated on the same dataset \cite{dataset}.
In addition to the proposed method, the pre-trained network $\mathcal{M}$ without the training using $\mathcal{D}$ was also evaluated (Pre-train) to investigate the performance improvement by the proposed training for improving the OSQA score.

Table~\ref{tbl:obj} summarizes the evaluated scores, where the proposed method achieved the state-of-the-art score for PESQ compared to the conventional methods. 
As the proposed method can be considered as an improved version of MetricGAN, the higher PESQ score indicates the effectiveness of the proposed method.
Although the proposed method did not outperform conventional methods on the other metrics, it is a straightforward result because the proposed method in this experiment was specialized to PESQ and did not take the other scores into account.
To improve these scores simultaneously, design of mixed-OSQA as in \cite{Koizumi_TASL_2018} should be performed.
However, since there is no OSQA which perfectly correlates with the sound quality, improving every OSQA score is not the essential goal for improving the actual subjective quality, at least for the current standards.
\subsection{Subjective evaluation}
To confirm whether the proposed method improved the actual subjective quality, we conducted a subjective experiment.
The proposed method was compared with SEGAN \cite{segan} and DF-Loss \cite{dfl} because speech samples of these methods are openly available in the web-page \cite{dfl_web}.
We selected 20 samples from Tranche 1--4 data (low SNR conditions) from the web-page.
The speech samples of the proposed method used in this test are also openly available\footnote{\url{https://miyazaki-lab.github.io/icassp2020_demo/}}.
Nine participants evaluated the sound quality of the output signals according to ITU-T P.835 \cite{P835}.
The participants listened to each test sample three times and evaluated the quality of only speech (S-MOS), only noise (N-MOS), and overall (G-MOS).
By evaluating S-MOS and N-MOS before evaluating G-MOS, the situation that only either one of the speech or noise affects the score of G-MOS was avoided.

Figure~\ref{fig:sub} shows the results of the subjective evaluation. 
The proposed method outperformed SEGAN in terms of all factors, and outperformed DF-Loss except N-MOS. 
In addition, statistically significant differences were observed in S-MOS and G-MOS between the proposed method and the others according to the paired one sided $t$-test ($p< 0.01$).
This result suggests that the proposed method improves not only objective metrics but also subjective quality.
\section{Conclusion}
In this study, we proposed the use of stabilization techniques to the function-approximation-based method for stably improving OSQA scores.
For stably training the auxiliary DNN approximating OSQA, we designed a new cost function (\ref{eq:OurlossD}) and adopted training techniques as described in Sec.\,\ref{sec:training_tech}.
Experiments showed that the proposed method
(i) was able to stably train the DNN,  
(ii) achieved the state-of-the-art PESQ score on the public dataset, and
(iii) obtained better subjective quality than the conventional methods.
Thus, we concluded that the proposed method is effective for (i) stabilizing the training of DNN-based speech enhancement for increasing OSQA score, and (ii) improving subjective quality of the enhanced signal.
%
\clearpage
\bibliographystyle{IEEEbib}

\end{document}